\documentclass[aps,pre,reprint,groupedaddress,amsmath,amssymb,amsfonts,floatfix]{revtex4-2}

\usepackage{float}
\usepackage{graphicx}
\usepackage[caption=false]{subfig}
\usepackage[colorlinks,linkcolor=blue,anchorcolor=blue,citecolor=blue,urlcolor=blue]{hyperref}

\begin{document}


\title{Higher-order evolutionary dynamics with game transitions}


\author{Yi-Duo Chen}

\author{Zhi-Xi Wu}

\author{Jian-Yue Guan}
\email[Corresponding author: ]{guanjy@lzu.edu.cn}

\affiliation{Lanzhou Center for Theoretical Physics, Key Laboratory of Theoretical Physics of Gansu Province, Key Laboratory of Quantum Theory and Applications of MoE, Gansu Provincial Research Center for Basic Disciplines of Quantum Physics, Lanzhou University, Lanzhou 730000, China}%


\date{\today}

\begin{abstract}

Higher-order interactions are prevalent in real-world complex systems and exert unique influences on system evolution that cannot be captured by pairwise interactions. We incorporate game transitions into the higher-order prisoner's dilemma game model, where these transitions consistently promote cooperation. Moreover, in systems with game transitions, the proportion of higher-order interactions has a dual impact, either enhancing the emergence and persistence of cooperation or facilitating invasions that promote defection within an otherwise cooperative system. Correspondingly, bistable states, consisting of mutual defection and either mutual cooperation or coexistence, are consistently identified in both theoretical analyses and simulation results. 
\end{abstract}


\maketitle



\section{\label{intro}Introduction}

In the natural world, higher-order interactions play a crucial role in the evolution of complex systems. Traditional graphs, which consist of edges connecting pairs of nodes, fail to capture the nonlinear interactions occurring among more than two individuals. In contrast, higher-order networks, such as hypergraphs and simplicial complexes, provide powerful tools for accurately modeling systems with higher-order interactions \cite{lambiotte2019networks,battiston2020Networks,majhi2022Dynamics,iacopini2024temporal,wang2024Epidemica,battiston2021physics,feng2024Hypernetwork,digaetano2024Percolation,courtney2016Generalized,boccaletti2023structure,kundu2022Higherorder,skardal2021Higherorder}. Research on consensus dynamics has demonstrated that nonlinear interaction functions reveal the higher-order properties of these systems \cite{neuhauser2020Multibody}. Other researchers demonstrated that higher-order components play a pivotal role in shaping contagion dynamics on hypergraphs \cite{kim2024Higherorder}. Besides, different higher-order network structures exhibit diverse mechanisms influenced by higher-order interactions, such as the opposing effects of promotion and inhibition on phase oscillator synchronization \cite{zhang2023Higherorder}. In addition to the examples listed, higher-order interactions exhibit many unique effects on system evolution \cite{wang2024Epidemica,battiston2020Networks,boccaletti2023structure}.

Research on the linear public goods game in hypergraphs has demonstrated the impact of group size and structural properties on the evolution of cooperation \cite{alvarez-rodriguez2021Evolutionary}. Moreover, the nonlinear effects of multi-individual interactions play a crucial role in the evolutionary dynamics of populations. Marginal utility represents a realistic form of nonlinear higher-order interactions in real-world systems and has been found to promote cooperation across various network structures in the public goods game model \cite{sheng2024Strategy}. A related study further demonstrated that increasing the degree of nonlinearity of public goods games enhances the synergistic promotion of cooperation \cite{zhu2024Evolutionary}. Recently, explosive transitions to cooperative states were observed as the higher-order properties of a network gradually strengthened in the higher-order prisoner's dilemma game model \cite{civilini2024Explosive}. These findings underscore the non-negligible influence of higher-order interactions in understanding the mechanisms driving the emergence and evolution of cooperation.

Studies on the coevolutionary game dynamics of strategies and environments have garnered increasing attention in recent years \cite{weitz2016oscillating,hauert2019Asymmetric,lin2019Spatial,tilman2020Evolutionary,chen2025Coevolutionary}. Game transitions (GT) capture the spatially heterogeneous environmental feedback in evolutionary games \cite{hilbe2018Evolution,su2019Evolutionary}. For instance, mutual cooperation improves the environment, leading to higher payoffs for individuals in this state, whereas defection degrades the environment, restricting individual payoffs. This mechanism is closely related to partner-fidelity feedback in evolutionary biology, as seen in the grass-endophyte mutualism \cite{su2019Evolutionary,bull1991Distinguishing,sachs2004Evolution,schardl1997Evolution,cheplick2009Ecology}. Notably, GT serves as an effective mechanism for enhancing the evolutionary competitiveness of cooperation, beyond the well-known five rules \cite{hilbe2018Evolution,su2019Evolutionary,nowak2006Five}. Further research has explored the effects of reputation mechanisms, reinforcement learning, environmental information, and state-dependent strategies on the evolution of cooperation in systems with game transitions \cite{feng2024evolutionary,huang2020Learning,kleshnina2023effect,wang2021Evolution}.

As demonstrated previously, higher-order interactions are widespread in the natural world \cite{wang2024Epidemica,battiston2020Networks,boccaletti2023structure}, suggesting that understanding feedback-driven evolutionary dynamics involving higher-order interactions is a crucial question that remains to be addressed. Furthermore, multi-individual feedback mechanisms have been empirically demonstrated across diverse biological systems, including reciprocal helping among cooperatively breeding birds \cite{Earl2025cryptic}, partner-reward feedback in mycorrhizal mutualism \cite{Toby2011Reciprocal}, and localized resource-mediated interactions between bacterioplankton and chitin-derived public goods \cite{Pollak2021Public}. Accordingly, we introduce game transitions into the higher-order prisoner's dilemma model. Our findings demonstrate that game transitions consistently promote the evolution of cooperation. Meanwhile, higher-order interactions can exert both facilitating and inhibiting effects on cooperative behaviors, with variations in the proportion of such interactions potentially leading to bistable states of cooperation (or coexistence) and defection. These phenomena depend sensitively on the intensity of the social dilemma and the disparity of transitions. The replicator dynamic approximation results generally align well with simulations, while also revealing more intricate transition phenomena in parameter space under specific conditions. Additionally, simulation results reveal more cooperative behaviors that are not captured by the theoretical approximation, particularly in cases where a greater payoff disparity between game states, along with a higher proportion of higher-order interactions, further enhances cooperation.

The remainder of the article is organized as follows. In Sec.~\ref{model}, we introduce the model and describe the MC simulation procedures in detail. Sec.~\ref{rd_ta} presents a theoretical analysis of the temporal evolution of cooperation based on replicator dynamics. Simulation and theoretical results under various scenarios of social dilemmas are presented and discussed in Secs.~\ref{rd_gm}--\ref{rd_zero_a}. In Sec.~\ref{gtpc}, we provide a simplified example of the imitation process to illustrate the mechanism by which game transitions promote cooperation. Finally, Sec.~\ref{conclusion} summarizes our main findings.

\section{\label{model}Model and simulations}

\begin{figure}
    \includegraphics[width=0.9\linewidth]{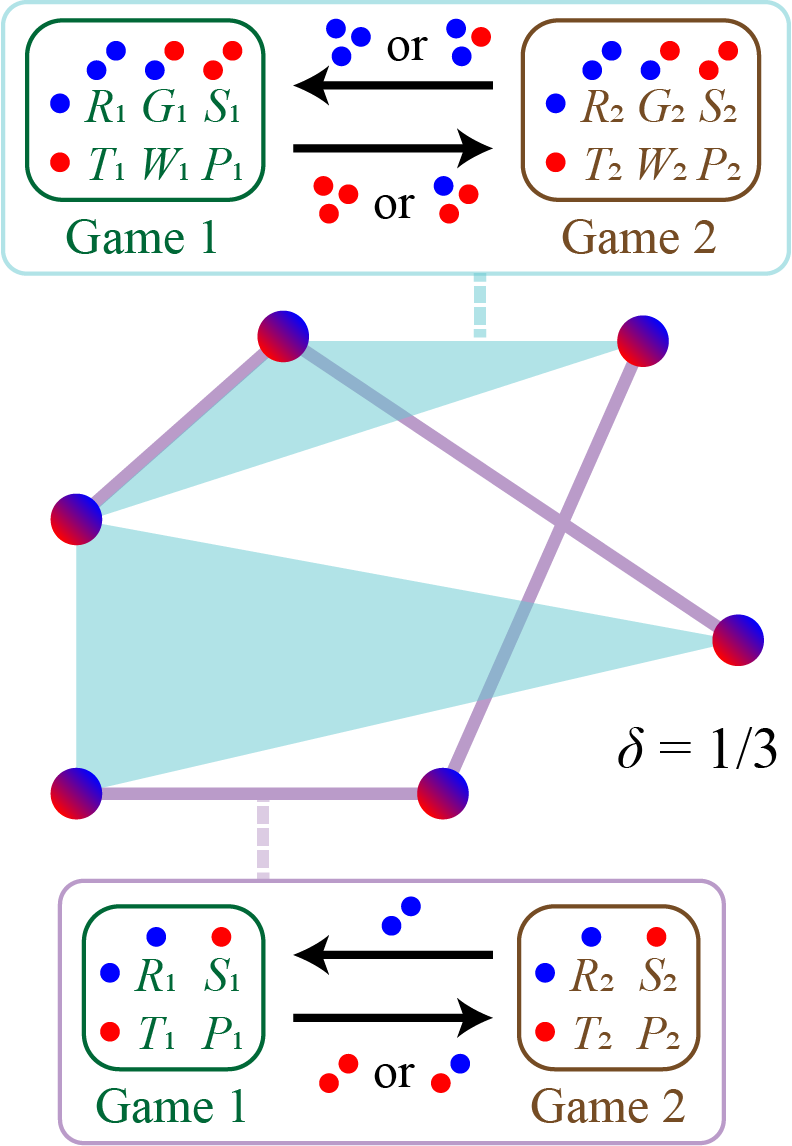}

  \caption{\label{fig_m}Interactions in the system for $\delta = 1/3$, $N=6$, and $L=6$. The red and blue gradient-color dots are nodes, the blue triangles are 3-hyperedges, and the purple lines are 2-edges, respectively. The blue (purple) rectangular box presents the game transitions on 3-hyperedges (2-edges). The blue (red) dots are cooperators (defectors). Individuals' payoff matrices change from game 1 ($\Pi_1^{(3)}$ and $\Pi_1^{(2)}$) to game 2 ($\Pi_2^{(3)}$ and $\Pi_2^{(2)}$) when there are less than two cooperators in the edges (hyperedges) and change from game 2 to game 1 otherwise. }    
  
\end{figure}

We examine the evolution of strategies on network structures that consist of both higher-order interactions and pairwise interactions \cite{civilini2024Explosive}. The network consists of $N$ nodes and $L$ edges. The parameter $\delta$ determines the fraction of 3-hyperedges (edges representing interactions among three individuals) relative to the total number of edges, while the remaining edges are standard 2-edges (pairwise interactions). Consequently, the network contains $\delta L$ 3-hyperedges and $(1-\delta)L$ 2-edges. In this framework, the network forms a uniform hypergraph of order $g=3$ when $\delta=1$ and a conventional random network when $\delta=0$ \cite{alvarez-rodriguez2021Evolutionary}. As illustrated in Fig.~\ref{fig_m}, for $\delta=1/3$, $N=6$, and $L=6$, the network comprises six nodes (represented as red and blue gradient-colored dots) connected by two 3-hyperedges (blue triangles) and four 2-edges (purple lines). Each edge is associated with a game state $\eta \in \{1, 2\}$, and all edges initially have the game state $\eta_{0} = 1$ in simulations.

Each individual $i$ in the system occupies a node and is assigned a strategy $s_i \in \{C, D\}$. The degree of node $i$ is given by $k = k_2 + k_3$, where $k_2$ represents the number of nodes connected to $i$ via pairwise edges, and $k_3$ denotes the number of 3-hyperedges in which $i$ is involved. Thus, the average degree of nodes in the system is $\langle k \rangle = \frac{2(1-\delta)L + 3\delta L}{N}$. Individuals engage in donation games within the edges and hyperedges to which they are connected. Consequently, the payoff matrix for the three-player game is  
\begin{equation}\label{pm_3}
  \Pi_\eta^{(3)} = 
  \begin{bmatrix}
    R_\eta & G_\eta & S_\eta \\
    T_\eta & W_\eta & P_\eta \\
  \end{bmatrix}, 
\end{equation}
where $\eta \in \{1, 2\}$ represents the game state of the edge where the game occurs. Here, $R_\eta$ denotes the reward for mutual cooperation, $S_\eta$ represents the sucker's payoff, $T_\eta$ denotes the temptation to defect, and $P_\eta$ is the punishment for mutual defection. Additionally, $G_\eta$ ($W_\eta$) represents the payoff for cooperation (defection) against one cooperator and one defector.  
Similarly, the payoff matrix for the two-player game is  
\begin{equation}\label{pm_2}
  \Pi_\eta^{(2)} = 
  \begin{bmatrix}
    R_\eta & S_\eta \\
    T_\eta & P_\eta \\
  \end{bmatrix}, 
\end{equation}
where the values of the elements remain consistent with those in Eq.~\eqref{pm_3}.  
In the donation game, cooperators incur a cost ($c$) to provide a benefit ($b$) to their opponents, whereas defectors contribute nothing. Thus, the elements are defined as follows: 
\begin{equation}
  \begin{aligned}
    R_\eta &= b-c- \Theta_\eta e, \\
    T_\eta &= b- \Theta_\eta e, \\
    G_\eta &= 0.5+a/4- \Theta_\eta e, \\
    W_\eta &= 0.5-a/4- \Theta_\eta e, \\
    S_\eta &= -c, \quad P_\eta=0,
  \end{aligned}
\end{equation}
where $e$ represents the payoff disparity between distinct game states; thus, $\Theta_\eta$ is defined as $\Theta_\eta = 0$ when $\eta = 1$, corresponding to cases in abundant environments, and $\Theta_\eta = 1$ when $\eta = 2$, corresponding to lower-payoff cases in damaged environments. 
And $a = 2(G_\eta - W_\eta) \in (-2, 2)$ controls the intensity of the three-player dilemma \cite{civilini2024Explosive,nash1950Equlibrium}. When $a > 0$ ($a < 0$), individuals receive higher (lower) payoffs for cooperation when their two opponents in the three-player game are a cooperator and a defector. The case $a = 0$ corresponds to a scenario in which both strategies yield equal payoffs. We also set $b > c > 0$ for the donation games considered in this study. 
Payoff matrix $\Pi_1^{(3)}$ ($\Pi_1^{(2)}$) represents a prisoner's dilemma game model of three (two) players \cite{civilini2024Explosive}. 
It is important to note that, depending on the value of $e$, game state 2 does not always correspond to a prisoner's dilemma game.

In each time step of the simulation, a randomly chosen individual $i$ has a probability $\epsilon = 2/N$ of randomly updating its strategy (mutation) to either cooperation ($C$) or defection ($D$), regardless of its previous strategy. In expectation, this results in the random appearance of one cooperator (defector) in a fully defective (cooperative) system. If individual $i$ does not update its strategy randomly, it selects a neighboring individual $j$ and imitates its strategy $s_j$ with probability: 
\begin{equation}\label{imitate_prob}
  P(s_j \to s_i)= \frac{1}{1+{\rm exp}[\omega(\pi_i-\pi_j)]},
\end{equation}
where $\omega$ represents the selection intensity, and $\pi_i$ ($\pi_j$) denotes the total payoff of individual $i$ ($j$).  
We adopt a local transition mechanism for game states, meaning that individuals actively modify their environment while participating in games \cite{su2019Evolutionary}. After imitation is determined, individuals $i$ and $j$ update the game states of all edges they are connected to, based on their own and their neighbors' strategies. To illustrate the cooperative repair effect on the local environment, we define game state 1 (the high-payoff state) as occurring when at least two cooperators exist within an edge (see Fig.~\ref{fig_m}). Consequently, edges with fewer than two cooperators contribute less to overall game value. We set the benefit that individuals in state 2 receive from cooperators within the same edge (when there is at least one cooperator among opponents) to be reduced by a value of $e$ compared to state 1, indicating that all cooperators lower their contributions to coplayers in a deteriorated environment.

We define $N$ time steps as one Monte Carlo (MC) step, ensuring that each individual has an equal chance to update its strategy once. To investigate the evolutionary dynamics, we measure the fraction of cooperators as follows:
\begin{equation}\label{eq_foc}
  x = \frac{N_C}{N},
\end{equation}
where $N_C$ is the number of cooperators in the system. Without loss of generality, we set $N=5000$, $\omega=1/\langle k \rangle$, $b=1.1$, $c=0.1$, and $\langle k \rangle=15$ in the Monte Carlo simulations. When performing the simulations, the total number of edges $L = L_2 + L_3$ is determined by the values of $N$, $\delta$ and $\langle k \rangle$. Specifically, we compute
\begin{equation}\label{eq_L2_3}
  L_3 = \frac{\delta \langle k \rangle N}{\delta + 2}, 
  \quad
  L_2 = \frac{(1 - \delta) \langle k \rangle N}{\delta + 2},
\end{equation}
where both $L_2$ and $L_3$ are rounded to integers for implementation. 
It is important to note that the initial fraction of cooperators is randomly selected to investigate the multiple stable states of the system. Subsequently, individual strategies are assigned according to the chosen probability.

\section{\label{result}Results and discussion}

\subsection{\label{rd_ta}Theoretical analysis}

We employ Replicator Dynamics to analyze the conditions under a well-mixed structure and an infinitely large system ($N \to \infty$) \cite{hofbauer1998Evolutionarya,traulsen2005Coevolutionarya,traulsen2006Coevolutionarya,traulsen2006Stochastica}. In this case, the mutation probability is given by $2/N \to 0$ as $N \to \infty$, which can be neglected. The temporal evolution of the fraction of cooperators is described by 
\begin{equation}\label{rd_eq_0}
  \dot{x}=x(1-x)(\pi_C - \pi_D),
\end{equation}
where $\pi_C$ ($\pi_D$) denotes the average payoff of cooperators (defectors). To simplify the game state transitions in the theoretical analysis, we assume that the game state of edges updates immediately following the strategy update of the associated individuals. An individual plays the game in state 1 when at least two cooperators are present on an edge (hyperedge). Consequently, the simplified payoff matrix for a three-player game is
\begin{equation}\label{simp_pm_3}
  \Pi^{(3)'} = 
  \begin{bmatrix}
    b-c & 0.5+a/4 & -c \\
    b & 0.5-a/4-e & 0 \\
  \end{bmatrix}, 
\end{equation}
and the payoff matrix for a two-player game is
\begin{equation}\label{simp_pm_2}
  \Pi^{(2)'} = 
  \begin{bmatrix}
    b-c & -c \\
    b-e & 0 \\
  \end{bmatrix}. 
\end{equation}
Furthermore, the average payoff disparity between cooperators and defectors is given by
\begin{equation}\label{payoff_diff}
  \begin{aligned}
    &\pi_C - \pi_D = (1-\delta)[x(e-c)+(1-x)(-c)]\\
  &\quad +\delta[x^2(-c)+2x(1-x)(e+a/2)+(1-x)^2(-c)]\\
  &=-\delta(a+2c+2e)x^2+[e+\delta(a+2c+e)]x-c. 
  \end{aligned}
\end{equation}
The temporal evolution of the fraction of cooperators is then expressed as
\begin{equation}\label{rd_eq}
  \dot{x}=x(1-x)(u_1 x^2 + u_2 x - c),
\end{equation}
where $u_1 = -\delta(a+2c+2e)$ and $u_2 = e+\delta(a+2c+e)$. The fixed points are given by $x_0^* = 0$, $x_1^* = 1$, and $x_{\pm}^* = \frac{-u_2 \pm \sqrt{\Delta}}{2u_1}$, where $\Delta = u_2^2 + 4u_1 c$. The stability of these fixed points is analyzed in Appendix~\ref{app_1}. 
By numerically integrating Eq.~\eqref{rd_eq} using the fourth-order Runge-Kutta method (hereafter referred to as the numerical results), we obtain the temporal evolution of the fraction of cooperators.


\begin{figure*}
  \includegraphics[width=\linewidth]{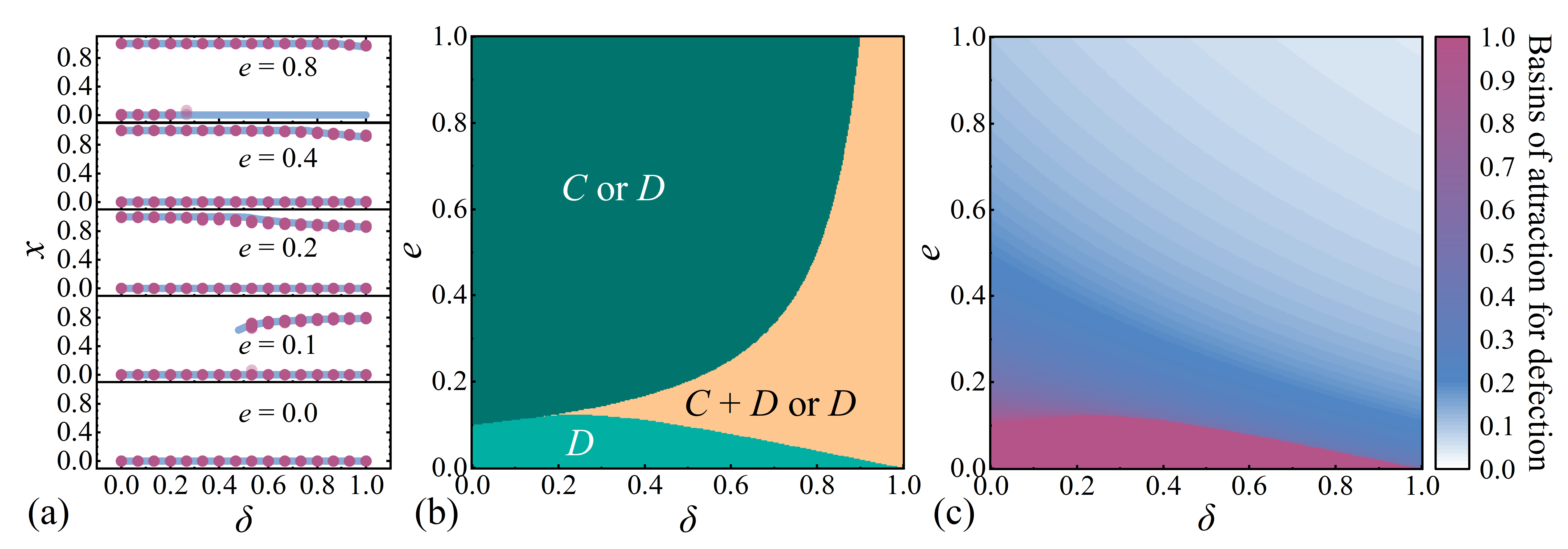}

\caption{\label{fig_hm_a02}The stability of the fraction of cooperators for $a=0.2$. (a) The fraction of cooperators as a function of $\delta$ for $e=0$, $0.1$, $0.2$, $0.4$, and $0.8$. The red points (60 \% transparency) represent results from Monte Carlo simulations once the system has stabilized (300 repeats with different random values of the fraction of cooperators for same other parameters), while the blue curves show the results obtained through numerical integration of Eq.~\eqref{rd_eq} using the RK4 method. (b) Stability of fixed points in parameter space of $\delta$ and $e$. (c) Basins of attraction for mutual defection states in parameter space of $\delta$ and $e$. }    

\end{figure*}


\begin{figure*}
  \includegraphics[width=\linewidth]{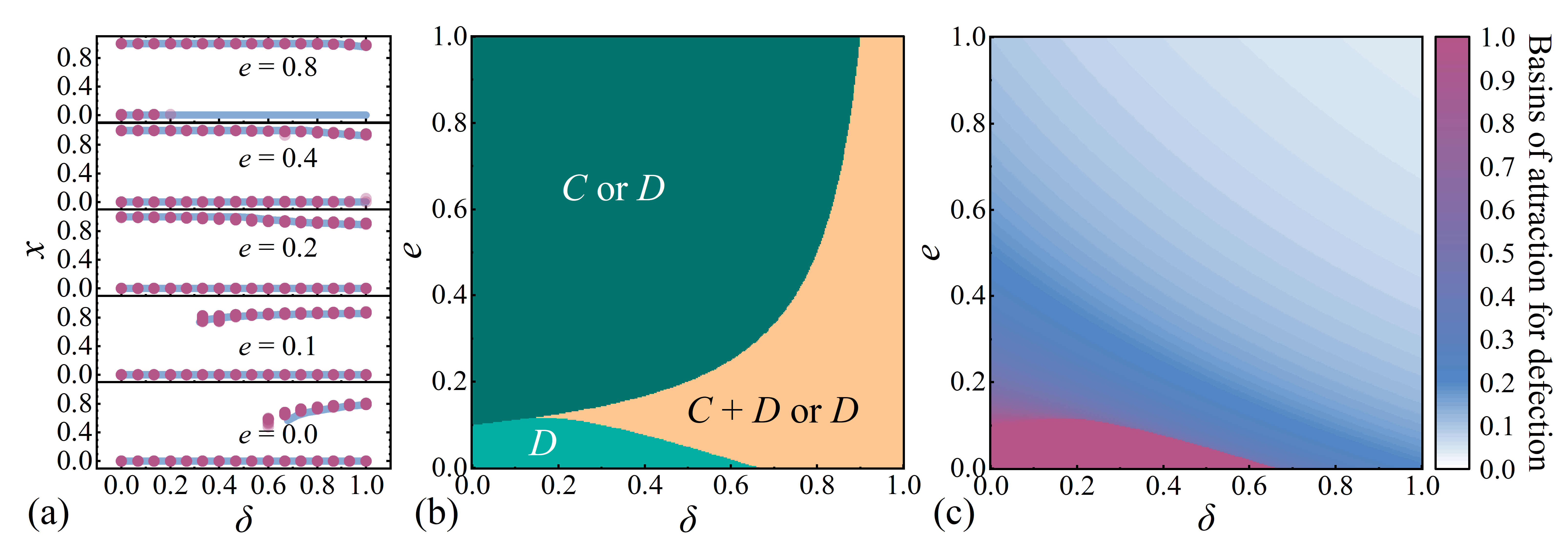}

\caption{\label{fig_hm_a04}The stability of the fraction of cooperators for $a=0.4$. (a) The fraction of cooperators as a function of $\delta$ for $e=0$, $0.1$, $0.2$, $0.4$, and $0.8$. The red points (60 \% transparency) represent results from Monte Carlo simulations once the system has stabilized (300 repeats with different random values of the fraction of cooperators for same other parameters), while the blue curves show the results obtained through numerical integration of Eq.~\eqref{rd_eq} using the RK4 method. (b) Stability of fixed points in parameter space of $\delta$ and $e$. (c) Basins of attraction for mutual defection states in parameter space of $\delta$ and $e$. }    

\end{figure*}


\begin{figure*}
  \includegraphics[width=\linewidth]{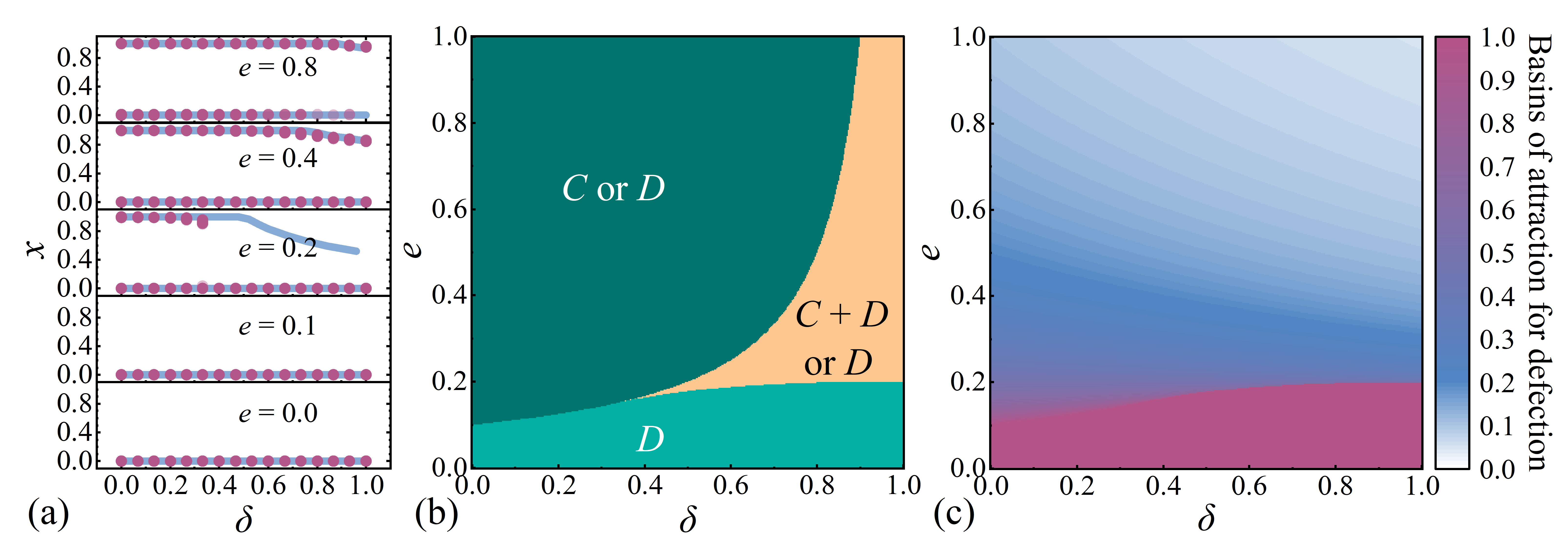}

\caption{\label{fig_hm_an02}The stability of the fraction of cooperators for $a=-0.2$. (a) The fraction of cooperators as a function of $\delta$ for $e=0$, $0.1$, $0.2$, $0.4$, and $0.8$. The red points (60 \% transparency) represent results from Monte Carlo simulations once the system has stabilized (300 repeats with different random values of the fraction of cooperators for same other parameters), while the blue curves show the results obtained through numerical integration of Eq.~\eqref{rd_eq} using the RK4 method. (b) Stability of fixed points in parameter space of $\delta$ and $e$. (c) Basins of attraction for mutual defection states in parameter space of $\delta$ and $e$. }    

\end{figure*}


\begin{figure*}
  \includegraphics[width=\linewidth]{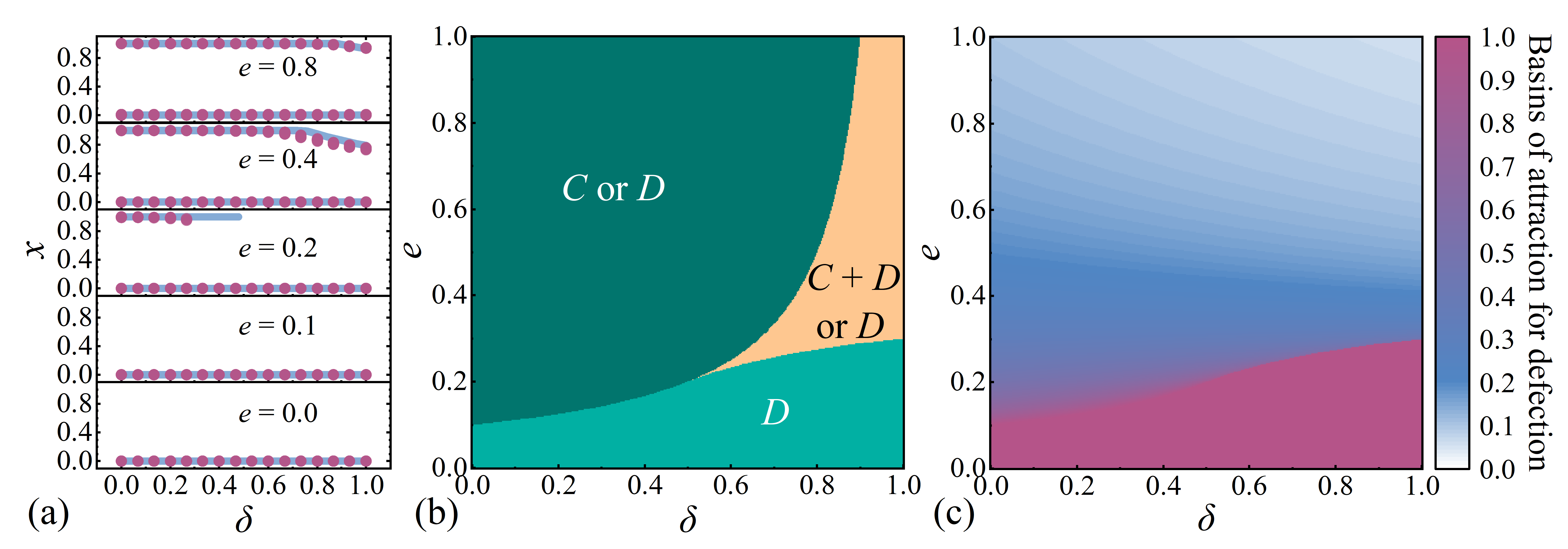}

\caption{\label{fig_hm_an04}The stability of the fraction of cooperators for $a=-0.4$. (a) The fraction of cooperators as a function of $\delta$ for $e=0$, $0.1$, $0.2$, $0.4$, and $0.8$. The red points (60 \% transparency) represent results from Monte Carlo simulations once the system has stabilized (300 repeats with different random values of the fraction of cooperators for same other parameters), while the blue curves show the results obtained through numerical integration of Eq.~\eqref{rd_eq} using the RK4 method. (b) Stability of fixed points in parameter space of $\delta$ and $e$. (c) Basins of attraction for mutual defection states in parameter space of $\delta$ and $e$. }    

\end{figure*}


\begin{figure*}
  \includegraphics[width=\linewidth]{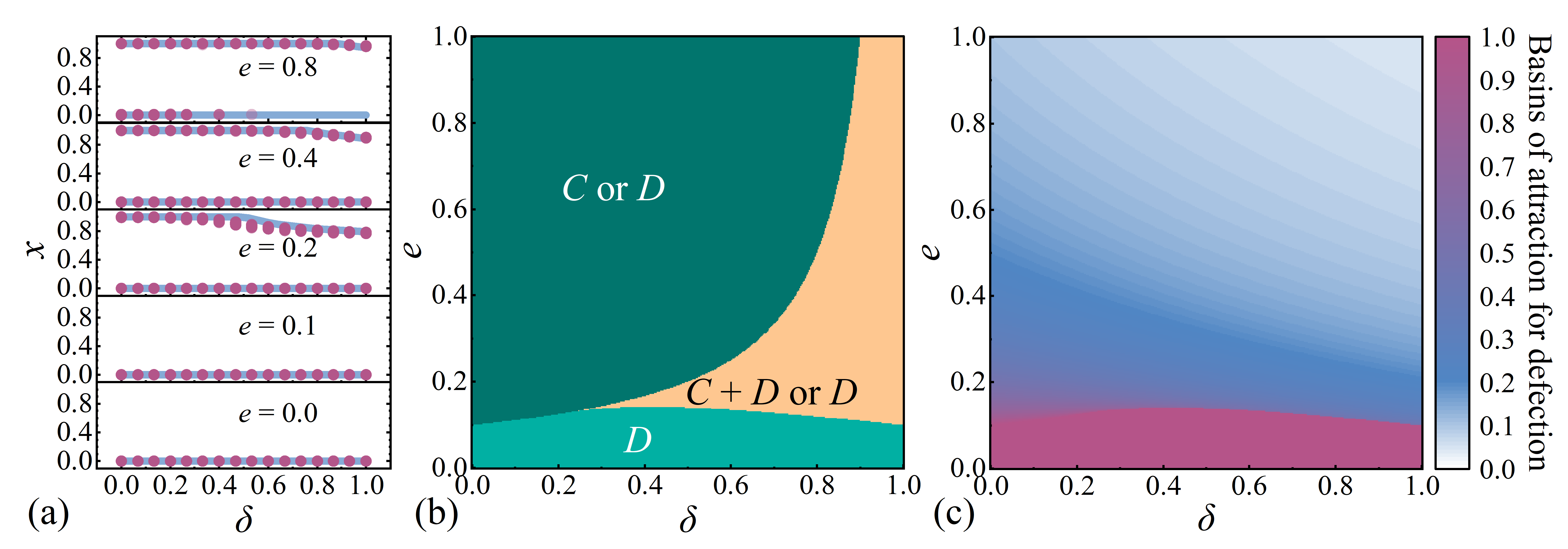}

\caption{\label{fig_hm_a00}The stability of the fraction of cooperators for $a=0$. (a) The fraction of cooperators as a function of $\delta$ for $e=0$, $0.1$, $0.2$, $0.4$, and $0.8$. The red points (60 \% transparency) represent results from Monte Carlo simulations once the system has stabilized (300 repeats with different random values of the fraction of cooperators for same other parameters), while the blue curves show the results obtained through numerical integration of Eq.~\eqref{rd_eq} using the RK4 method. (b) Stability of fixed points in parameter space of $\delta$ and $e$. (c) Basins of attraction for mutual defection states in parameter space of $\delta$ and $e$. }    

\end{figure*}



\subsection{\label{rd_gm}General model details and results}

As set in the model, $a$ controls the intensity of the three-player prisoner's dilemma in game state 1 by influencing the Nash equilibria (NE) \cite{civilini2024Explosive,nash1950Equlibrium}. When $a > 0$ (with $b=1.1$ and $c=0.1$), the NEs of the three-player prisoner's dilemma game are $(C, C, D)$ and $(D, D, D)$, where individual order does not matter [i.e., $(C, C, D)$ is equivalent to $(D, C, C)$]. These cases are further discussed in Sec.~\ref{rd_positive_a}. For $a < 0$, the only NE of the rigorous three-player prisoner's dilemma is $(D, D, D)$, which we analyze in Sec.~\ref{rd_negative_a}. The case of $a=0$ is a special scenario where any individual can randomly choose strategies without affecting payoffs when its opponents in the hyperedge include both a cooperator and a defector. This scenario is examined in Sec.~\ref{rd_zero_a}. 

Game transitions become trivial when $e = 0$, in which case our model closely resembles that of Ref.~\cite{civilini2024Explosive}. The primary distinction is that while they employed the quasistationary (QS) method \cite{deoliveira2005How,meleard2012Quasistationary,sander2016Sampling,zhou2010Evolutionary} to prevent the system from reaching an absorbing state and analyzed phase transitions, we introduce mutations as noise in our model. Figs.~\ref{fig_hm_a02}(a), \ref{fig_hm_a04}(a), \ref{fig_hm_an02}(a), \ref{fig_hm_an04}(a), and \ref{fig_hm_a00}(a) confirm that our results align with theirs \cite{civilini2024Explosive}.  

Figs.~\ref{fig_hm_a02}(b), \ref{fig_hm_a04}(b), \ref{fig_hm_an02}(b), \ref{fig_hm_an04}(b), and \ref{fig_hm_a00}(b) illustrate the stability of the fixed points within the parameter space defined by $e$ and $\delta$. The labels in these figures are as follows: ``$C$ or $D$'' (dark green area) indicates a bistable state of mutual cooperation and mutual defection, ``$D$'' (light green area) represents a stable state of mutual defection, and ``$C+D$ or $D$'' (yellow area) denotes a bistable state where cooperators and defectors coexist alongside mutual defection. Figs.~\ref{fig_hm_a02}(c), \ref{fig_hm_a04}(c), \ref{fig_hm_an02}(c), \ref{fig_hm_an04}(c), and \ref{fig_hm_a00}(c) illustrate the basins of attraction for mutual defection states, where the color of each point represents the proportion of defection states for specific conditions in parameter space. 

It is important to note again that the game matrices for state 2 do not always correspond to prisoner's dilemma games for the range of $e$ values considered. When $e \gg 0$, the payoff matrix parameters become predominantly negative, rendering these cases less meaningful for discussion. Therefore, our investigation is limited to cases where $e$ ranges from 0 to 1. 

For the convenience of explanation, we refer to any state where the fraction of cooperators (such as in the mutual cooperation state or coexistence state) exceeds the fraction caused solely by mutations (approximately $1/N$) as the cooperation state. 
As shown in Fig.~\ref{fig_hm_a02}(a), when $e=0$ and $a=0.2$, the MC simulation results (red points) indicate that almost no cooperators persist in the system. Unlike the numerical results (blue lines), where cooperators completely vanish in the mutual defection state, the presence of mutations with a low probability ($2/N$) allows a small number of cooperators to persist in simulations.


\subsection{\label{rd_positive_a}The weak three-player prisoner's dilemma}

A positive value of $a$ leads to a weak dilemma in the three-player game. When $a$ is sufficiently great and $e=0$ [see Fig.~\ref{fig_hm_a04}(a)], a bistable state of cooperation and defection emerges for greater values of $\delta$, which aligns well with Ref.~\cite{civilini2024Explosive}. This suggests that a weaker dilemma (greater $a$) in higher-order interactions enables cooperators to resist defectors' invasions and maintain a stable fraction of cooperators, $x^*>0$. Conversely, when the initial fraction of cooperators $x$ is too low, cooperators fail to invade defective regions.  

Increasing $e$ to a sufficiently great value introduces a critical value of $\delta$ for $a>0$, beyond which greater $\delta$ drives the system into a bistable state of cooperation and defection [\textit{e.g.}, $e=0.1$ in Figs.~\ref{fig_hm_a02}(a) and \ref{fig_hm_a04}(a)]. Thus, when the initial fraction of cooperators is great enough, the system remains in the cooperation state. Otherwise, the fraction of cooperators declines rapidly, leading to a defection state. Furthermore, a weaker dilemma lowers the threshold for cooperation emergence; the critical $\delta$ value is lower for $a=0.4$ than for $a=0.2$.  

A further increase in $e$ reveals different phenomena. As shown in Figs.~\ref{fig_hm_a02}(a) and \ref{fig_hm_a04}(a), when $e=0.2$ and $e=0.4$, the system maintains a bistable state of cooperation and defection over the entire range $\delta \in [0,1]$. However, beyond a critical $\delta$ value, the stable fraction of cooperators gradually decreases. This suggests that defectors can exploit cooperators in hyperedges of three individuals without degrading the environment enough to lower payoffs. This occurs because hyperedges tend to remain in state 1, which yields higher payoffs when either only one defector is present or defectors are entirely absent. Greater $\delta$ increases the likelihood of this phenomenon.  

In the previously presented results, numerical solutions align well with MC simulations. However, due to limitations of the replicator dynamics, such as assuming a well-mixed structure, a linear relationship between payoffs and fitness, and the absence of pattern formation and stochastic dynamics, discrepancies arise for greater $e$. The Fermi imitation probability used in our simulations allows for irrational individual choices. The stochastic dynamics in this context provide a pathway out of mutual defection as the dilemma is weakened by GT. Moreover, clusters of cooperators or defectors in structured populations are well known for their ability to prevent invasions by individuals adopting the opposite strategy. A notable example is $e=0.8$, where numerical results predict that the system should maintain bistable states of cooperation and defection for $\delta \in [0,1]$. In contrast, MC simulations indicate that the cooperation state becomes the only stable state for great value of $\delta$, with lower critical values for greater $a$ [see Figs.~\ref{fig_hm_a02}(a) and \ref{fig_hm_a04}(a)]. 

Moreover, theoretical analysis (see Appendix~\ref{app_1} for details) of the stability of the replicator equation [Eq.~\eqref{rd_eq}] reveals more complex dynamical phenomena [see Figs.~\ref{fig_hm_a02}(b) and (c), and \ref{fig_hm_a04}(b) and (c)]. In the parameter space defined by $e$ and $\delta$, three stable states emerge (detailed label explanations are provided in Sec.~\ref{rd_gm}). The variation of $a$ significantly affects the regions occupied by these three states. In general, a greater $a$ (i.e., a weaker dilemma) promotes cooperation by shrinking the basins of attraction for defection states and facilitating transitions to more cooperative states. Additionally, as indicated by theoretical analysis, numerical simulations, and MC results, game transitions (i.e., increasing the payoff disparity $e$) consistently enhance cooperation. The impact of $\delta$ itself is evident in Figs.~\ref{fig_hm_a02}(b) and (c), and \ref{fig_hm_a04}(b) and (c), where it corresponds to a horizontal line. These results also suggest that higher-order interactions can either enhance the competitiveness of cooperators or facilitate the exploitation of cooperators by defectors within cooperative neighborhoods. Notably, a greater $\delta$ always leads to fewer defective basins of attraction in bistable states.


\subsection{\label{rd_negative_a}The rigorous three-player prisoner's dilemma}

In simulations of the rigorous three-player prisoner's dilemma ($a<0$), the results contrast with those observed in the weak dilemma. As shown in Figs.~\ref{fig_hm_an02}(a) and \ref{fig_hm_an04}(a), game transitions are not sufficiently effective in promoting cooperation when the value of $e$ is low. Moreover, when $e=0.2$, systems with a low proportion (i.e., low $\delta$) of higher-order interactions exhibit a bistable state of cooperation and defection. The fundamental mechanism by which game transitions promote cooperation is the enhancement of cooperators' payoff advantages within edges of state 1. However, when the dilemma is rigorous (i.e., negative values of $a$), game transitions are less effective at fostering cooperation in the three-player game than in the two-player game, despite sharing four identical payoff matrix elements ($R_\eta$, $S_\eta$, $T_\eta$, and $P_\eta$). Furthermore, the effectiveness of game transitions diminishes as the dilemma becomes more severe. This implies that systems with lower values of $a$ have a lower critical $\delta$, marking the transition from a bistable state of cooperation and defection to a state of mutual defection. 

When game transitions are effective enough to generally promote cooperation, the system remains in a bistable state of cooperation and defection. As shown in Figs.~\ref{fig_hm_an02}(a) and \ref{fig_hm_an04}(a), when $e=0.4$ and $0.8$, both simulation and numerical results consistently indicate bistable cooperative states for $\delta \in [0,1]$. In Figs.~\ref{fig_hm_an02}(a) and \ref{fig_hm_an04}(a), numerical results exhibit significant deviations only for $e$ values near $0.2$, due to reasons distinct from those previously discussed. When $\delta$ is sufficiently great, defectors can stably persist in the system by exploiting the payoffs of cooperators without degrading the game states and overall payoffs. Additionally, the presence of a single stable cooperation state in Fig.~\ref{fig_hm_an02}(a) for $e=0.8$ and $\delta=1$ in simulations further demonstrates that higher-order interactions continue to promote cooperation in this case. Although the coexistence ratio of cooperators has decreased, the mutual defection state are less reached with the increase of $\delta$. 

As shown in Figs.~\ref{fig_hm_an02}(b) and (c), and \ref{fig_hm_an04}(b) and (c), three stable states exist within the parameter space defined by $e$ and $\delta$, with label meanings explained in Sec.~\ref{rd_gm}. An increase in $a$ significantly influences the regions occupied by these three states and reduces the basins of attraction for defection states. An increase in the payoff disparity $e$ generally promotes cooperation. Moreover, when all other parameters are held constant while $\delta$ alone increases (impact of $\delta$ is also reflected as a horizontal line in these figures), the system generally transitions to less cooperative states. However, this also leads to fewer basins of attraction for defection states. Thus, the percentage of higher-order interactions has dual effects on cooperation in a rigorous dilemma.


\subsection{\label{rd_zero_a}Medium intensity of the three-player dilemma}

Individuals receive identical payoffs for either cooperation or defection when interacting with both a cooperator and a defector in the three-player game with $a = 0$. Thus, we refer to this as the medium-intensity case. Notably, we do not observe transitions from bistable states to mutual defection or vice versa for $\delta \in [0, 1]$ when other parameters remain constant [see Fig.~\ref{fig_hm_a00}(a)]. As shown in Fig.~\ref{fig_hm_a00}, other results closely resemble those obtained for different values of $a$. Game transitions consistently promote cooperation, whereas a high proportion of higher-order interactions may provide defectors with a competitive advantage for persistence. Consequently, for $a=0$, higher-order interactions can either facilitate or inhibit (also appear in the impact of boundaries of states in parameter space and alter the basins of attraction for the defection state) cooperation in all results, depending on the specific parameter conditions.

Since our simulations employ a random initial fraction of cooperators, the probability of the system remaining in the defection state gradually approaches zero as $\delta$ increases for great value of $e$. Consequently, when $\delta > 0.3$, only a few data points fall into the defection state in Fig.~\ref{fig_hm_a00} for $e=0.8$. This phenomenon occurs in all simulation results with large values of $e$, but this particular plot provides the clearest illustration. The defection states gradually vanish as the value of $\delta$ increases greater than $0.2$.

\subsection{\label{gtpc}An intuitive explanation of the role of game transitions}

\begin{figure}
  \includegraphics[width=\linewidth]{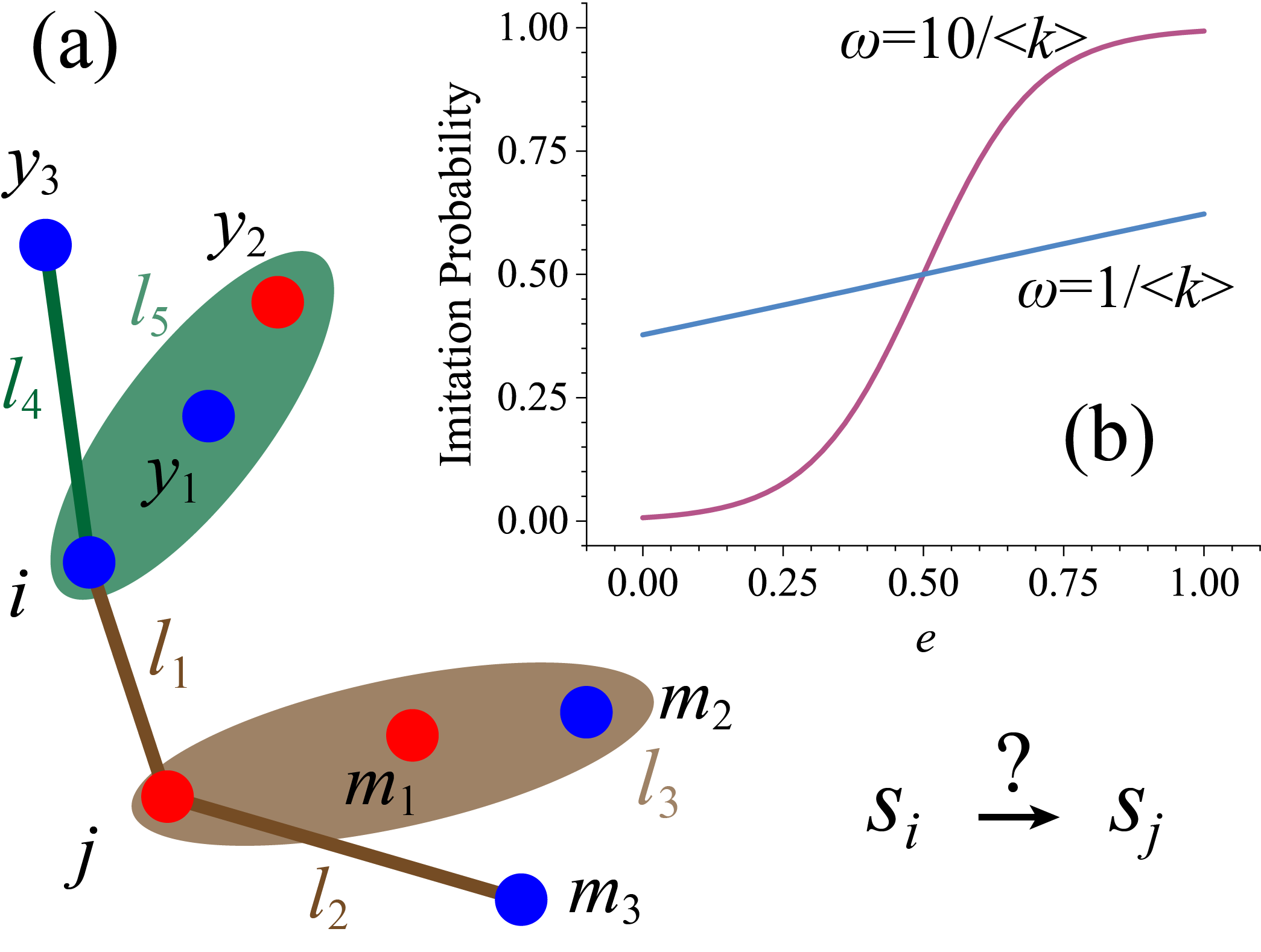}

\caption{\label{fig_gt}Imitation process in a simplified scenario. (a) Local network structure in the case where the model player is $i$ and the focal player is $j$, with $\langle k \rangle = 3$. Nodes $j$ ($i$), $y_1$ ($m_1$), $y_2$ ($m_2$), and $y_3$ ($m_3$) are neighbors of node $i$ ($j$). Connection structures for nodes other than $i$ and $j$ are not shown here. Lines and ovals represent 2-edges and 3-hyperedges, respectively. Blue (red) nodes represent cooperators (defectors), and green (brown) edges indicate game state 1 (2). (b) The imitation probability of player $j$ adopting player $i$'s strategy ($C$) as a function of the payoff disparity $e$ for $a = -0.4$. The blue curve corresponds to $\omega = 1/\langle k \rangle$, which is the value used throughout this paper, while the red curve, corresponding to $\omega = 10/\langle k \rangle$, serves as a reference. }
\end{figure}

As mentioned earlier, game transitions enhance the competitiveness of cooperators and promote the evolution of cooperation under great values of the payoff disparity $e$. To illustrate this principle, we consider a simplified imitation scenario. As shown in Fig.~\ref{fig_gt}(a), individual $i$ (cooperator) is the model player and $j$ (defector) is the focal player. We impose identical conditions on the neighbors of individuals $i$ and $j$, excluding the connection between them. However, the states of the edges are not identical due to the influence of their own strategies. Individual $i$ ($j$) obtains its payoff by participating in games along edges $l_1$, $l_4$ ($l_2$), and $l_5$ ($l_3$). Thus, the payoff of individual $i$ is
\begin{equation}
  \Pi_i = S_2 + R_1 + G_1,
\end{equation}
and the payoff of individual $j$ is
\begin{equation}
  \Pi_j = T_2 + T_2 + W_2,
\end{equation}
so the payoff difference between individuals $j$ and $i$ is
\begin{equation}
  \Pi_j - \Pi_i = 1.5 - 3e,
\end{equation}
and the probability of individual $j$ adopting $i$'s strategy is given by
\begin{equation}
  P(s_i \to s_j)= \frac{1}{1+{\rm exp}[\omega(1.5 - 3e)]}.
\end{equation}
As shown in Fig.~\ref{fig_gt}(b), the imitation probability increases with the payoff disparity $e$. In the absence of game transitions ($e=0$), individual $j$ adopts $i$'s strategy with an extremely low probability. However, for $e > 0.5$, $j$ is much more likely to adopt $i$'s strategy. The blue curve corresponds to the weak selection case ($\omega = 1/\langle k \rangle$) used throughout this paper, while the red curve represents a stronger selection intensity ($\omega = 10/\langle k \rangle$), where the impact of $e$ is more pronounced. Similarly, in more complex scenarios, game transitions generally promote cooperation by enhancing the relative payoff advantages of cooperators.

The promotion effect of game transitions on the evolution of cooperation is essentially equivalent to Parrondo's Paradox~\cite{parrondo1998efficiency,dinis2008optimal,Parrondo01032004,parrondo2000new,harmer1999losing}. Specifically, the prisoner's dilemma game in state 1 and a lower-payoff game in state 2, each of which alone does not promote cooperation, can jointly enhance cooperative behavior when combined. Even in cases with a positive value of $a$, where the three-player game in state 1 alone promotes cooperation, the combination of the two games (positive $e$) yields an even stronger cooperative outcome (see figures \ref{fig_hm_a02} and \ref{fig_hm_a04}).

Recent studies in reinforcement learning that incorporate environmental information have elucidated the mechanisms underlying the emergence of cooperation and trust \cite{Zheng2024decoding,Zheng2024Evolution}. Therefore, beyond the imitation mechanism considered in this work, the influence of learning paradigms on higher-order game systems with environmental feedback remains an open and important direction for future research.

\section{\label{conclusion}Conclusion}

In this work, we introduce game transitions into a higher-order prisoner's dilemma model. The system is represented as a hypergraph consisting of 2-edges and 3-hyperedges, with a parameter $\delta$ controlling the proportion of hyperedges (i.e., higher-order interactions). When $\delta = 0$, the system reduces to a random graph, whereas $\delta = 1$ corresponds to a hypergraph composed exclusively of 3-node hyperedges. 

We demonstrate that game transitions (with payoff disparity $e>0$) consistently promote cooperation. However, higher-order interactions can both facilitate and inhibit the evolution of cooperation. Depending on $\delta$, the system may exhibit bistable states of cooperation (mutual cooperation or coexistence state) and defection, or transition to a mutual defection state, even under the same values of $e$ and $a$ (controls the intensity of three-player dilemma). Theoretical analyses also demonstrate that greater $\delta$ lead to fewer basins of attraction for defection state in bistable states. Simulation results reveal similar regularities, with greater $\delta$ values further reducing the persistence of defectors within predominantly cooperative populations, and, in some cases, leading to a stable state of mutual cooperation. A positive $a$ weakens the dilemma, generally fostering cooperation in hypergraphs, except in coexistence states for $\delta \to 1$, where a small fraction of defectors persist within cooperative systems. Conversely, a negative $a$ intensifies the dilemma, leading to predominantly defective outcomes. Nonetheless, higher-order interactions continue to exert a dual influence on cooperation in this case. 

Furthermore, analytical results reveal more intricate boundary structures in the parameter space of $e$ and $\delta$, predicting certain transitions that do not emerge in simulations. These discrepancies arise because theoretical analyses cannot fully account for structured finite populations, the nonlinear relationship between payoff and fitness, local transitions, and stochastic dynamics in simulations.

\begin{acknowledgments}
  This work was supported by the National Natural Science Foundation of China (Grants Nos.~11475074, 12247101 and 12375032), the Fundamental Research Funds for the Central Universities (Grant Nos.~lzujbky-2025-it50, lzujbky-2024-11, lzujbky-2023-ey02 and lzujbky-2024-jdzx06), the Natural Science Foundation of Gansu Province (No.~22JR5RA389), and the 111 Center under Grant No.~B20063. 
  This work was partly supported by Longyuan-Youth-Talent Project of Gansu Province. 
  We gratefully acknowledge the referees for their diligent efforts and insightful comments, which have substantially enhanced the quality and clarity of this manuscript. 

\end{acknowledgments}

\section*{Data availability}
The data that support the findings of this article are openly available \cite{data_avai}.

\appendix

\begin{figure}
  \includegraphics[width=\linewidth]{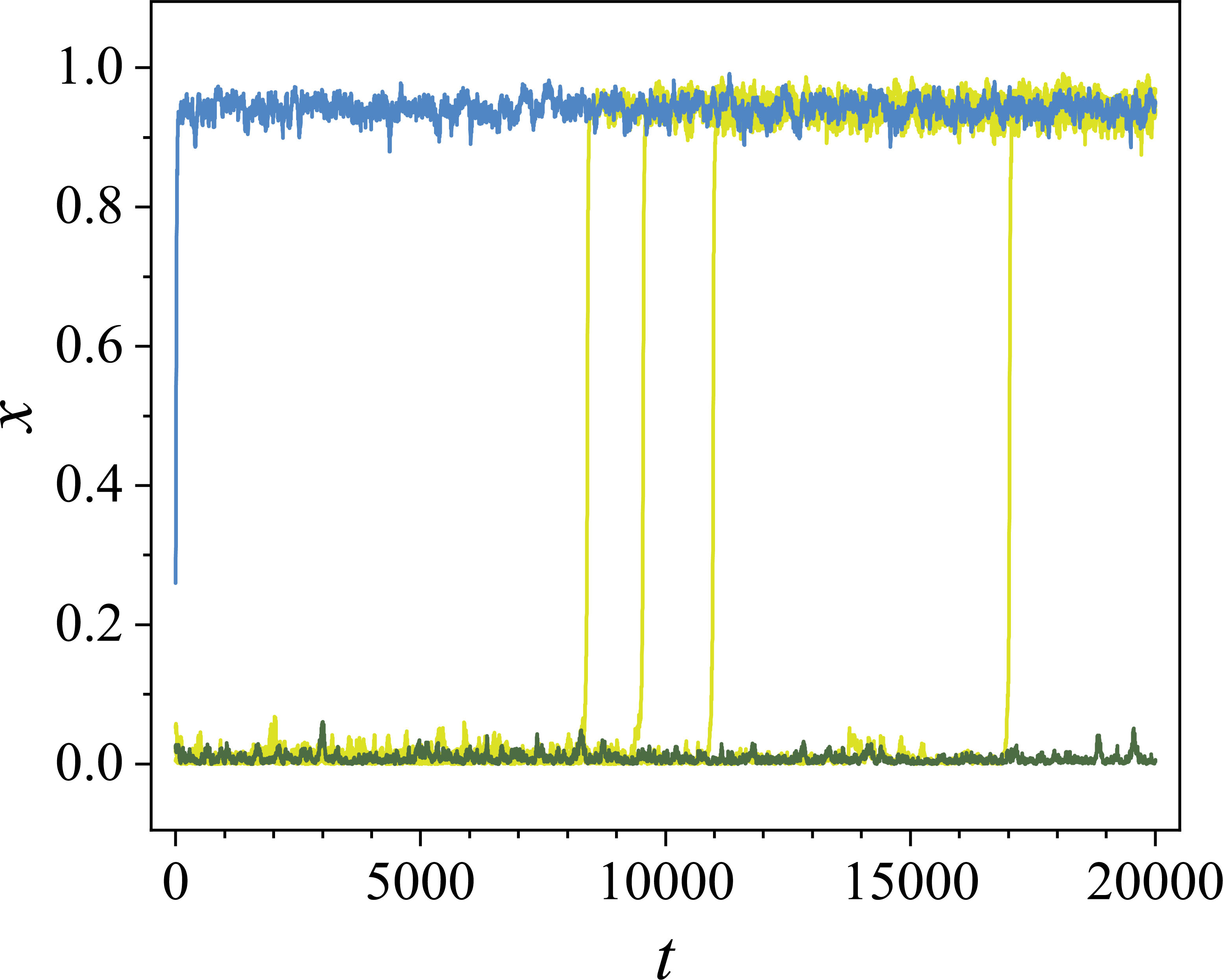}

\caption{\label{fig_ts_d}The fraction of cooperators as a function of time for $a=0.4$, $e=0.4$, and $\delta=1.0$ with different values of $x_0$. The blue (green) line represents the system evolving to a state with more (fewer) cooperators. Yellow lines indicate cases where the system undergoes dramatic state transitions between the two aforementioned states during evolution. }    

\end{figure}

\begin{figure}
  \includegraphics[width=0.8\linewidth]{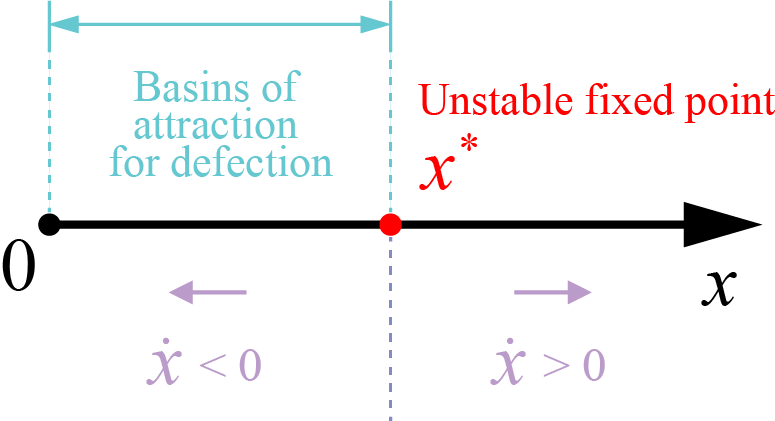}

\caption{\label{fig_ba_d}The first unstable fixed point $x^*>0$ next to $x_0^* = 0$ marks the boundary of the basins of attraction for defection. The red point represents $x^*$. The blue area represents the basins of attraction for defection. Light purple arrows mark the direction of the temporal evolution of the fraction of cooperators $x$ according to Eq.~\eqref{rd_eq}. }    

\end{figure}

\begin{figure}
  \includegraphics[width=\linewidth]{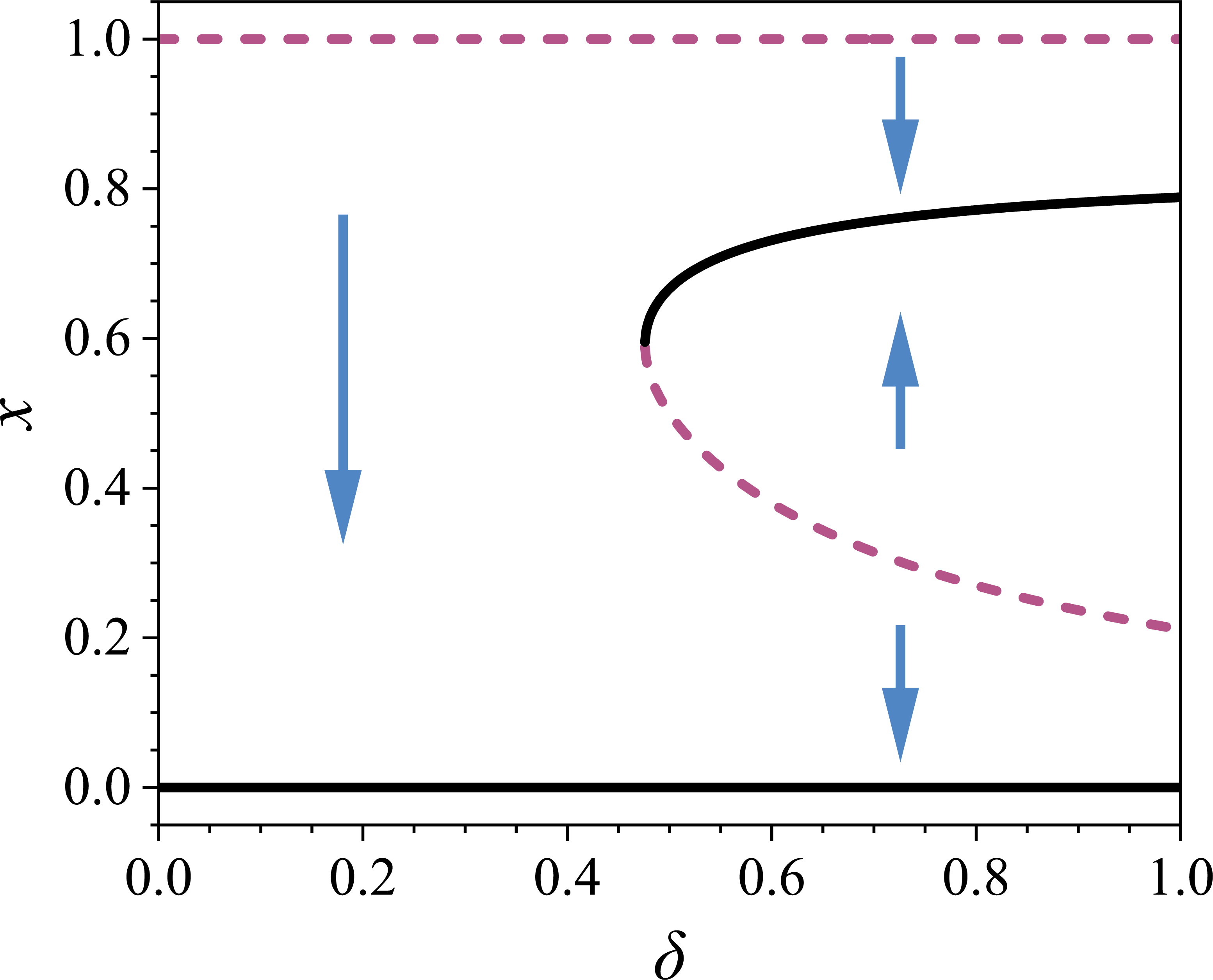}

\caption{\label{fig_bif_s_us}The fixed points of the fraction of cooperators $x$ as a function of $\delta$ for $a=0.2$ and $e=0.1$. Black curves represent stable solutions, red dashed curves indicate unstable solutions, and blue arrows illustrate the temporal evolution of $x$ according to Eq.~\eqref{rd_eq}. }    

\end{figure}

\section{\label{app_1}Stability of fixed points of replicator dynamics}

For our model, we should always set $b>c>0$. 
Based on Eq.~\eqref{rd_eq}, the stability of the fixed-points can be analyzed through the sign of the function $f(x) = u_1 x^2 + u_2 x - c$. Since the relation $x(1-x) > 0$ holds for $x \in (0, 1)$, the function $f(x)$ either describes a quadratic parabola when $u_1 \neq 0$ or a straight line when $u_1 = 0$. 
Without loss of generality, we classify fixed point stability into two categories in the following discussion: stable and unstable. This is because semistable fixed points cannot maintain stability in a system subject to noise, which we explicitly consider.

The stability conditions for the fixed points are as follows: 
(1) Bistable state of mutual defection ($x_0^*$) and mutual cooperation ($x_1^*$): \{$u_1=0$ and $u_2>c$\}, \{$u_1>0$, $\Delta>0$, $x_-^* \leq 0$ and $0<x_+^*<1$\}, and \{$u_1<0$, $\Delta>0$, $x_-^* \geq 1$ and $0<x_+^*<1$\}; 
(2) Only one stable state of mutual defection ($x_0^*$): \{$u_1=0$ and $u_2 \leq c$\}, \{$u_1<0$ and $\Delta < 0$\}, \{$u_1>0$, $\Delta>0$, $x_-^* \leq 0$ and $x_+^* \geq 1$\},  \{$u_1<0$, $\Delta \geq 0$, $x_-^* \leq 0$ and $x_+^* \leq 0$\}, \{$u_1<0$, $\Delta \geq 0$, $x_-^* \geq 1$ and $x_+^* \geq 1$\}, and \{$u_1<0$, $\Delta=0$\}; 
(3) Only one stable state of mutual cooperation ($x_1^*$): \{$u_1>0$ and $\Delta<0$\}, \{$u_1>0$, $\Delta=0$\}, \{$u_1>0$, $\Delta>0$, $x_-^* <0$ and $x_+^* \leq 0$\}, \{$u_1>0$, $\Delta>0$, $x_-^* \geq 1$ and $x_+^* >1$\}, and \{$u_1<0$, $\Delta>0$, $x_-^* \geq 1$ and $x_+^* \leq 0$\}; 
(4) Only one stable state of the coexist state of cooperators and defectors ($x_{\pm}^*$): \{$u_1>0$, $\Delta>0$, $0 < x_-^* < 1$ and $x_+^* \geq 1$\} and \{$u_1<0$, $\Delta>0$, $0 < x_-^* < 1$ and $x_+^* \leq 0$\}; 
(5) Bistable state of mutual cooperation ($x_1^*$) and coexist state of cooperators and defectors ($x_{\pm}^*$): \{$u_1>0$, $\Delta>0$ and $0<x_-^* <x_+^* < 1$\}; 
(6) Bistable state of mutual defection ($x_0^*$) and coexist state of cooperators and defectors ($x_{\pm}^*$): \{$u_1<0$, $\Delta>0$, $0<x_+^* <x_-^* < 1$\}. 

The results presented in this paper illustrate only a subset of these cases. The model's dynamics under different parameter settings can be derived based on the conditions we outline here.

\section{\label{app_2}Supplemental information and discussion}

As shown in Fig.~\ref{fig_ts_d}, the fraction of cooperators does not remain stable over long-term evolution due to the mutations of strategy updates among individuals. The system can easily transition to another stable state and remain there for an unpredictable duration. The blue and green lines represent examples where the system remains in a single stable state from the beginning, whereas the yellow lines illustrate cases where the system transitions between stable states after a uncertain period. These phenomena are observed across different parameter values. Consequently, to ensure a reliable calculation of the average fraction of cooperators in the simulations, we exclude data points that have not reached stability within the time window $t = 15\thinspace 000$ to $t = 20\thinspace 000$. 

Moreover, as discussed in Appendix \ref{app_1}, Eq.~\eqref{rd_eq} exhibits semistable fixed points. To eliminate their influence in numerical integrations, we apply both positive and negative perturbations to $x$ once its value stabilizes and then reintegrate until a stable state is reached. Consequently, only stable fixed points appear in the plots. 

In the two bistable cases discussed, the unstable fixed point, where the minimum fraction of cooperators exceeds that of the mutual defection state, marks the boundary of the basins of attraction for the defection state (see Figs.~\ref{fig_ba_d} and \ref{fig_bif_s_us}). The red dashed curve in Fig.~\ref{fig_bif_s_us}, located beneath the black curve corresponding to the cooperative state $x > 0$, delineates the boundary of the basins of attraction for the defection state (i.e., the red point in Fig.~\ref{fig_ba_d}). Figs.~\ref{fig_hm_a02}(c), \ref{fig_hm_a04}(c), \ref{fig_hm_an02}(c), \ref{fig_hm_an04}(c), and \ref{fig_hm_a00}(c) are plotted based on this phenomenon.

%

\end{document}